\documentclass[twocolumn,showpacs,preprintnumbers,amsmath,amssymb]{revtex4}
\usepackage{amsmath}

\begin{document}

\tolerance=5000

\def\pp{{\, \mid \hskip -1.5mm =}}
\def\cL{{\cal L}}
\def\be{\begin{equation}}
\def\ee{\end{equation}}
\def\bea{\begin{eqnarray}}
\def\eea{\end{eqnarray}}
\def\beq{\begin{eqnarray}}
\def\eeq{\end{eqnarray}}
\def\tr{{\rm tr}\, }
\def\nn{\nonumber \\}
\def\e{{\rm e}}

\title{Dark Energy and Modified Gravities}

\author{Shin'ichi Nojiri\footnote{E-mail: 
snojiri@yukawa.kyoto-u.ac.jp, nojiri@nda.ac.jp}}
\address{Department of Applied Physics,
National Defence Academy,
Hashirimizu Yokosuka 239-8686, JAPAN}

\begin{abstract}
We review the several models of the dark energy, which may generate 
the accelerated expansion of the present universe. 
We also discuss the the Big Rip singularity, which may occur when the equation of the state parameter 
$w$ is less than $-1$. We show that the quantum correction would be very important near the 
singularity.  
\end{abstract}


\maketitle

\section{Introduction}

By the recent observations of the universe, we have found that the universe is undergoing a phase of 
{\it accelerated} expansion at the present epoch from about 
$5\times 10^{10}$ years ago and the universe is {\it flat}.  

The accelerated expansion of the universe has been established by the observation of the 
type Ia Supernovae (SNIa) \cite{SN}. This type of  supernovae has an abosrption spectrum of Silicon, 
which helps us to distinguish the supernovae from other types. An important thing is that the 
intrinsic luminosities of SNIa are almost uniform and their absolute maginitude is almost $-20$, from 
which we can find the distance between the earth and the supernovae. From the distance, we can  
find how old the SNe are. We can also the speed of the SNe from the redshift. Combining the distance and 
the redshift, the accelerated expansion of the universe has been established. 

The flatness of the universe can be found from the observation of the 
anisotropies of Cosmic Microwave Background (CMB) observed by the baloons, 
the BOOMERANG project \cite{BOOMERANG} and the MAXIMA-1 project \cite{MAXIMA}. 
The spectrum of the CMB has a characteristic structure and the CMB was produced in the 
early universe, that is, when the universe becomes transparent or the mean free path of the light 
becomes infinite. Then the CMB has propagated very long distance, about 10 billion light years and 
if the univers is curved, the structure of the spectrum should be deformed. The deformation has, 
however, not observed, which tells that the universe should be flat. 

If the universe (the spacial part) is flat, we may assume the metric of the flat FRW universe 
in the following form: 
\be
\label{I}
ds^2= -dt^2 + a(t)^2\sum_{i=1,2,3} \left(dx^i\right)^2
\ee
Here $a$ is called a scale factor.  
If $\dot a > 0$, $\ddot a>0$, the universe is expanding and accelerating. 
Related with the accelerated expansion, we often use a parameter $w$, 
which is called as ``equation of state'' parameter. 
The present data tell
\be
\label{II}
w\equiv \frac{p}{\rho}\sim -1\ .
\ee
Here $p$ is the pressure and $\rho$ is the energy density of the universe. 
We may explain the relation between $w$ and the accelerated expansion later but 
if $w<-\frac{1}{3}$, the universe is accelerating, which Eq.(\ref{II}) tells. As $w$ is negative, if 
the energy density $\rho$ is positive as usual,  the pressure $p$ should be negative. 

In case of the radiation, we have $w=\frac{1}{3}$ since the trace of energy-momentum tensor 
vanishes: $T=-\rho + 3p=0$. Usual matters, like baryon, can be regarded as dust with $w=0$ ($p=0$). 
The cosmological term can gives $w=-1$. It is, however unnatural if the cosmological term, which may 
be suggested from (\ref{II}), is a part 
of the fundamental gravity. The cosmological constant should be given by $\Lambda\sim H^2$, 
where $H$ is the Hubble parameter, $H\equiv \frac{\dot a}{a}$. Since the Hubble parameter is given by
$H\sim$ 70 kms$^{-1}$Mpc$^{-1}$ $\sim$ $10^{-33}$ eV, the scale of the cosmological constant is 
unrealistically very small compared with the fundamental scale of the gravity, 
the Planck scale  $10^{19}$ GeV $=$ $10^{28}$ eV. 

On the other hand, since the universe is flat, 
the energy density of the universe should be almost critical density : 10$^{-29}$ g/cm$^3$. 
The density of the non-relativistic matter (baryons, dark matter) is, however, almost $\frac{1}{3}$ of the 
critical density. This tells that 70 $\%$ of the energy density of the universe should come from 
the matter (?) with negative $w$. We call the matter (?) as dark energy. Conversely, 
the density of the dark energy is almost same order with the density of the usual matter. 
Then maybe it would be more natural to consider that the dark energy would be a kind of matter, 
rather than that it would come from the fundamental cosmological term in the gravity. 

Now we explain the relation between the accelerated expansion and $w$. 
The energy conservation law $0=\nabla^\mu T_{\mu\nu}$ has the following form in the FRW metric (\ref{I}): 
\be
\label{III}
0=\dot \rho + 3H\left(\rho + p\right)=\dot \rho + 3(1+w)H\rho\ .
\ee
In case that $w$ is a constant, since $H=\frac{\dot a}{a}$, we can integrate (\ref{III}) :
\be
\label{IV}
\rho=\rho_0 a^{-3(1+w)}\ .
\ee
The (1st) FRW equation, which is the $(t,t)$ component of the Einstein equation, is given by
\be
\label{V}
\frac{6}{\kappa^2}H^2 =\rho \ .
\ee
Here $\kappa^2 = 16\pi G$ and $G$ is the gravitational constant. 
Then if $w\neq -1$, the scale factor $a$ can be found by integrating (\ref{V}) by using (\ref{IV}): 
\be
\label{VI}
a=a_0 t^\frac{2}{3(w+1)}\ .
\ee
The case $w=-1$ corresponds to deSitter universe:  
\be
\label{VII}
a=a_0\e^{\kappa t\sqrt{\frac{\rho_0}{6}}} \ .
\ee
Then if $w<-\frac{1}{3}$, that is, $\frac{2}{3(w+1)}>1$, the universe is accelerating, which can be found 
more directly, by combining 2nd FRW equation, 
\be
\label{VIII}
\frac{\ddot a}{a}=-\frac{\kappa^2}{12}\left(\rho + 3p\right)
=-\frac{\kappa^2}{12}\left(1+3w\right)\rho \ ,
\ee
which tells $\ddot a>0$ if $w<-\frac{1}{3}$. 

If $-1<w<-\frac{1}{3}$, the universe is expanding and accelerating and the corresponding matter is 
called as ``quintessence'' \cite{quintessence}. If $w<-1$, the universe is accelerating but shrinking 
in (\ref{VI}) but if we change the direction of time as $t\to t_s - t$, the universe is expanding 
(and accelerating). The corresponding matter is called as ``phantom'' \cite{phantom}. 

As clear from (\ref{VI}) after replacing $t\to t_s - t$, when $w<-1$, there is a singularity at $t=t_s$, 
which is called the Big Rip singularity. 
We will mention more about the singularity later.
The present data of the universe seems to tell that $w$ might be less than $-1$. In the viewpoint of the 
quantum field theory, however, this is rather strange. As an example, we consider the scalar field 
$\phi$ with the potential $V$. If we assume the FRW metric and $\phi$ only depends on the time 
coordinate, the enrgy density $\rho$ and the pressure $p$ are given by
\be
\label{IX}
\rho=\frac{1}{2}{\dot \phi}^2 + V(\phi)\ ,\quad 
p=\frac{1}{2}{\dot \phi}^2 - V(\phi)
\ee
Then if $V(\phi)>\frac{1}{2}{\dot \phi}^2$,  we have negative $w$ but the value of $w$ should be 
greater than $-1$. In order that $w<-1$, we should change the sign of the kinetic energy: 
\be
\label{X}
\rho=-\frac{1}{2}{\dot \phi}^2 + V(\phi)\ ,\quad 
p=-\frac{1}{2}{\dot \phi}^2 - V(\phi)\ ,
\ee
which tells that the phantom might be ghost with negative norm, which should be a serious problem in 
the quantum field theory \footnote{
The instability of the vacuum due to the quantum effect in the theories with negative kinetic energy
has been investigated in \cite{cline}.}. 
Conversely, if surely $w<-1$, there should be a new physics.

\section{Scalar with Exponential Potential}

In this section, as a most simple model, we consider the scalar field theory with the exponential potential. 
This model has been investigated from a very long time ago, say in \cite{expo}. 
The action of the model is given by
\bea
\label{XI}
S&=&\int d^4 x \sqrt{-g} \left( {1 \over \kappa^2}R - {\gamma \over 2}
\partial_\mu \phi \partial^\mu \phi - V(\phi) \right)\ ,\nn
V(\phi)&=&V_0 \e^{-{2\phi \over \phi_0}}
\eea
Usually $\gamma>0$, but for phantom, we may have $\gamma<0$. 
We assume (flat) FRW universe (\ref{I}) and $\phi$ only depend on time coordinate' $t$. Then the $\phi$-equation 
fo motion and the FRW equation has the following form:
\bea
\label{XII}
&& 0=-\gamma \left(\frac{d^2 \phi}{dt^2} + 3H \frac{d\phi}{dt}\right) - V'(\phi)\ \ ,\nn
&& {6 \over \kappa^2}H^2=\rho_\phi={\gamma \over 2}\left(\frac{d\phi}{dt}\right)^2 + V(\phi)
\eea
If we assume $\phi\propto \ln t$, $h\propto t^{-1}$, we can solve the equations in (\ref{XII}) very easily : 
\bea
\label{XIII}
&& \phi=\phi_0\ln \left|\frac{t}{t_1}\right|\ ,\quad 
H=\frac{\gamma \kappa^2\phi_0^2}{4 t}\ , \nn
&& t_1^2 \equiv -\frac{\gamma \phi_0^2 \left(1 - \frac{3\gamma \kappa^2\phi_0^2}{4}\right)}{2V_0}\ .
\eea
Then the scalae factor $a$ is given by
\be
\label{XIV}
a=a_0 \left|\frac{t}{t_1}\right|^{\frac{\gamma \kappa^2\phi_0^2}{4}}\ .
\ee
Since $a=a_0 t^\frac{2}{3(w+1)}$ for general $w$, we find 
\be
\label{XV}
w=-1 + \frac{8}{3\gamma \kappa^2\phi_0^2}
\ee
Then if $\gamma<0$, surely phantom appears. 

A very interesting point is that the general solutions of the scalar model with exponential potential can 
be found. In case $\gamma>0$ case \cite{Russo}, if we assume 
\bea
\label{XVI}
&& a=\e^{v+u \over
3}\ ,\quad \phi ={2(v-u) \over \sqrt{3\gamma}}\ ,\nn
&& d\tau = dt \sqrt{3V_0 \over 8}\e^{-{2(v-u) \over \phi_0\sqrt{3\gamma}}}\ ,
\eea
the Hamiltonian constraint and other equations have the following form:
\be
\label{XVII}
{dv \over d\tau}{du \over d\tau} =1
\ee
and
\be
\label{XVIIb} 
 {d^2 {\cal U} \over d\tau^2}=\left(1-{\bar \alpha}^2\right){\cal U}\ ,\quad 
{d^2 {\cal V} \over d\tau^2}=\left(1-{\bar \alpha}^2\right){\cal V}
\ee
Here
\be
\label{XVIII}
 {\cal V}\equiv \e^{(1 - {\bar \alpha})v}\ ,\quad
{\cal U}\equiv \e^{(1+{\bar \alpha})u}\ , \quad 
{\bar \alpha}\equiv {2 \over \kappa\phi_0\sqrt{3\gamma}}
\ee
The equations in (\ref{XVII}) can be solved very easily and when $\bar\alpha^2<1$, the behavior 
at $t\to +\infty$ coincides with that of the previous solution (\ref{XIII}), (\ref{XIV}) and $w$ has the 
following form:
\be
\label{XIX}
w=-1 + \bar\alpha^2
\ee
Then if $\bar\alpha^2<\frac{2}{3}$, the universe is expanding and accelerating (quintessence). 

The case $\gamma<0$ has been also exactly solved in \cite{ENO}, instead of (\ref{XVI}), by using a 
complex variable $z$, we assume
\bea
\label{XX}
&& a=\e^{z + z^*\over 3}\ , \quad \phi=-{2i\left(z - z^*\right) \over\sqrt{-3\gamma}}\ ,\nn
&& d\tau =  \pm dt \sqrt{3V_0 \over 8}\e^{-{2i(z-z^*) \over \phi_0\sqrt{-3\gamma}}} \ ,\nn
&& \tilde{\bar \alpha}\equiv {4 \over \kappa\phi_0\sqrt{-3\gamma}}\ ,\quad 
{\cal Z}\equiv \e^{(1 - i \tilde{\bar \alpha})z}
\eea
Then the Hamiltonian constraint and other equations have the following form:
\be
\label{XXI}
{dz \over d\tau}{dz^* \over d\tau}=1 \ , \quad 
{d^2 {\cal Z} \over d\tau^2}=\left(1 + \tilde {\bar \alpha}^2\right){\cal Z}
\ee
The above equations can be also solved easily. 
When $t\to 0$ (if we replace $t\to t_s - t$, $t\to t_s$), the behaviour corresponds to the one 
in the previous solution ad we find $w$ is given by
\be
\label{XXII}
w=-1 - {\tilde{\bar\alpha}}^2<-1 \ ,
\ee
which surely corresponds to the phantom. 

\section{A Modification of Exponential Potential Model}

Since the models with negative norm would not be consistent with quantum field theory as the theory may be 
non-unitary. We consider a modification of the model based on \cite{ENO}.

We may start with Jordan (string) frame action, which is the Brans-Dicke type \cite{BD}, 
\bea
\label{XXIII}
S&=&{1 \over \kappa^2}\int d^dx
\sqrt{-g}\e^{\alpha\phi}\left(R - {\gamma \over 2}
\partial_\mu \phi \partial^\mu \phi - V(\phi)\right) \nn
&& + \int d^dx \sqrt{-g}\left( -{1 \over 2}\partial_\mu \chi
\partial^\mu \chi - U(\chi)\right) \ .
\eea
Here $\alpha$ and $\gamma$ are constant parameters and $\gamma$ can be negative. 
The scalar field $\chi$ represents the  matter. Since $\chi$ does not interact directly with $\phi$, the 
equivalence principle would not be violated although 
the effective gravitational constant $\kappa\e^{-{\alpha \phi \over 2}}$ depends on $\phi$.

By using the scale transformation : $g_{\mu\nu}=\e^{-{2\alpha \over d-2}\phi}g_{E\,\mu\nu}$,  
the Jordan frame action (\ref{XXIII}) is transformed in the Einstein frame one:
\bea
\label{XXIV}
S&=&{1 \over \kappa^2}\int d^dx \sqrt{-g_E}\left(R_E - \left({(d-1)\alpha^2 \over d-2} \right.\right. \nn
&& \left.\left. +{\gamma \over
2}\right)g_E^{\mu\nu} \partial_\mu \phi \partial_\nu \phi - \e^{-{2\alpha \over d-2}\phi}V(\phi)\right) \nn
&& + \int d^dx \sqrt{-g_E}\left( -{\e^{-\alpha \phi} \over 2} g_E^{\mu\nu}\partial_\mu \chi
\partial_\nu \chi \right. \nn
&& \left. - \e^{-{d\alpha \over d-2}\phi}U(\chi)\right) \ .
\eea
Even if $\gamma<0$, when 
\be
\label{XXV}
 {(d-1)\alpha^2 \over d-2}+{\gamma \over 2}>0
\ee
the kinetic energy becomes positive. 
Especially when $d=4$ and $\gamma=-1$, we have $\alpha^2>\frac{1}{3}$. Then 
the sign of the kinetic energy depends on the frame.

When $d=4$, $\chi=0$ and $V(\phi)$ is exponential type, we can solve the system in a way similar to the 
last section. If we define,
\be
\label{XXVI}
\varphi\equiv \phi\sqrt{\alpha^2 + {\gamma \over 3}}\ ,\quad 
\tilde V(\varphi) \equiv \e^{-\alpha\phi}V(\phi)
=V_0 \e^{-2{\varphi \over \varphi_0}} 
\ee
and assume the FRW metric in Einstein frame : 
\be
\label{XXVII}
ds_E^2 = -dt_E^2 + a_E\left(t_E\right)^2\sum_{i=1,2,3}\left(dx^i\right)^2\ ,
\ee
we obtain the following solution
\bea
\label{XXVIIb}
&& a_E=a_{E0}\left({t_E \over t_{E0}}\right)^{{3 \over 4}\varphi_0^2}\ ,\quad 
\varphi =\varphi_0 \ln {t_E \over t_{E0}}\ ,\nn
&& t_{E0}\equiv \varphi_0 \sqrt{{1 \over V_0} \left({27 \over8}\varphi_0^2 - {3 \over 2}\right)}\ .
\eea
For the FRW metric in the original Jordan frame in (\ref{I}), we have
\bea
\label{XXVIII}
&& dt=\e^{-{\alpha \over 2}\phi}dt_E ={t_{E0}^{\beta \varphi_0 \over 2} \over 1 - {\beta\varphi_0 \over 2}}
d\left( t_E^{1- {\beta \varphi_0 \over 2}}\right) \ ,\nn
&& \beta\equiv {\alpha \over \sqrt{\alpha^2 + {\gamma \over 3}}}\ .
\eea
Then we obtain
\bea
\label{XXIX}
&& a=\e^{- {\alpha\phi \over 2}}a_E =a_{E0}\left({t \over t_0}\right)^{{3 \over 4}\varphi_0^2 - \beta\varphi_0
\over 1 - {\beta \varphi_0 \over 2}}\ ,\nn
&& \phi={\beta \varphi_0 \over \alpha \left(1 -{\beta \varphi_0 \over 2}\right)} \ln {t \over t_0} 
\ ,\quad t_0\equiv {t_{E0} \over 1 - 
{\beta \varphi_0 \over 2}}\ .
\eea
Then the condition for the acceleration is given by
\be
\label{XXX}
{{3 \over 4}\varphi_0^2 - {\varphi_0 \alpha \over 2\sqrt{\alpha^2 + {\gamma
\over 3}}} \over 1-{\varphi_0 \alpha \over 2\sqrt{\alpha^2 +{\gamma \over 3}}}}>1\ ,
\ee
and $w$ is given by
\bea
\label{XXX1}
w &=& -1 + {{2 \over 3}\left(1-{\varphi_0 \alpha \over 2\sqrt{\alpha^2 + {\gamma \over 3}}} \right) \over 
{3 \over 4}\varphi_0^2 - {\varphi_0 \alpha \over 2\sqrt{\alpha^2 + {\gamma \over 3}}}} \nn
&=&-{\varphi_0\left(2\beta - 9\varphi_0\right) + 8 \over3\left(2\beta - 3\varphi_0\right) \varphi_0}
\eea
Since the denominator in the above expression can vanish, by properly choosing the parameters $\alpha$, 
$\varphi_0$, ($\gamma$), we can put $w$ in an arbitrary value. 
Especially if ${\varphi_0 \alpha \over\sqrt{\alpha^2 + {\gamma \over 3}}}=2$, we have $w=-1$ and 
when $\varphi_0^2>\frac{4}{3}$ we find
\be
\label{XXX2} 
\begin{array}{lll}
w<-1 &\mbox{if}& {3 \over 2}\varphi_0^2 > {\varphi_0 \alpha \over \sqrt{\alpha^2 + {\gamma \over 3}}} > 2 \\
w>-1 &\mbox{if}& {\varphi_0 \alpha \over \sqrt{\alpha^2 + {\gamma \over 3}}}
> {3 \over 2}\varphi_0^2
\ \mbox{or}\ {\varphi_0 \alpha \over \sqrt{\alpha^2 + {\gamma \over 3}}} < 2 \ .
\end{array}
\ee
On the other hand, when $\varphi_0^2<{4 \over 3} $, 
\be
\label{XXX3}
\begin{array}{lll}
w<-1 &\mbox{if}& 2 > {\varphi_0 \alpha \over \sqrt{\alpha^2 + {\gamma \over 3}}}
> {3 \over 2}\varphi_0^2 \\
w>-1 &\mbox{if}& {\varphi_0 \alpha \over \sqrt{\alpha^2 + {\gamma \over 3}}}
< {3 \over 2}\varphi_0^2
\ \mbox{or}\ {\varphi_0 \alpha \over \sqrt{\alpha^2 + {\gamma \over 3}}}> 2 \ .
\end{array}
\ee
Even if $\gamma>0$, we can have $w<-1$. For example, if
\be
\label{XXX4}
\varphi_0=4\ ,\quad {\gamma \over \alpha^2}=3\  \left(\beta={1 \over \sqrt{2}}\right) \ ,
\ee
we have $w$, which is less than $-1$:
\be
\label{XXX5}
w=- {11\sqrt{2} + 203 \over 213}=-1.025...<-1\ .
\ee

\section{$1/R$-model and Generalization}

In \cite{CDTT}, a modification of Einstein gravity has been proposed. 
In this section, we consider the model and its generalization (see also \cite{ABF,NO1Rqc}). 

The action of the model in \cite{CDTT} include the inverse power of the curvature: 
\be
\label{XXX6}
S={1 \over \kappa^2}\int d^4 x \sqrt{-g} \left(R - {\mu^4 \over R}\right)\ ,
\ee
When curvature is small $\left(|R|\ll \mu^2\right)$, we have $a\propto t^2$ and $w=-\frac{2}{3}$. 
This model may be called as ``c-essence'' (``c'' expresses the curvature) \cite{Chiba}. 

We may consider the generalization of the model as follows,
\be
\label{XXX7}
S={1 \over \kappa^2}\int d^4 x \sqrt{-g} f(R)\ .
\ee
Here $f(R)$ can be an arbitrary function. 
By introducing the auxilliary fields $A$, $B$, we may rewrite the action (\ref{XXX7}) in the following form: 
\be
\label{XXX8}
S={1 \over \kappa^2}\int d^4 x \sqrt{-g} \left\{B\left(R-A\right) + f(A)\right\}\ .
\ee
By the variation of $B$, we obtain  $A=R$. By substituting this equation into the above action (\ref{XXX8}), 
we obtain the original action. If we consider the variation of $A$ first, we obtain $B=f'(A)$, which 
can be solved as $A=g(B)$. Then we obtain 
\be
\label{XXX9}
S={1 \over \kappa^2}\int d^4 x \sqrt{-g} \left\{B\left(R-g(B)\right) 
+ f\left(g(B)\right)\right\}\ .
\ee
Instead of $A$, we may delete $B$ as follows,
\be
\label{XXX10}
S={1 \over \kappa^2}\int d^4 x \sqrt{-g} \left\{f'(A)\left(R-A\right) + f(A)\right\}\ ,
\ee
which can be regarded as the Jordan frame action. 
By using the scale transformation $g_{\mu\nu}\to \e^\sigma g_{\mu\nu}$ with $\sigma = -\ln f'(A)$, 
we obtain the Einstein frame action : 
\bea
\label{XXX11}
\lefteqn{S_E=\left. {1 \over \kappa^2}\int d^4 x \sqrt{-g} \right\{ R }\nn
&& \left. - {3 \over 2}\left({f''(A) \over f'(A)}\right)^2 
g^{\rho\sigma}\partial_\rho A \partial_\sigma A - {A \over f'(A)} 
+ {f(A) \over f'(A)^2}\right\} \nn
&&={1 \over \kappa^2}\int d^4 x \sqrt{-g} \left( R - {3 \over 2}g^{\rho\sigma}
\partial_\rho \sigma \partial_\sigma \sigma - V(\sigma)\right)\ , \\
\lefteqn{V(\sigma)= \e^\sigma g\left(\e^{-\sigma}\right) - \e^{2\sigma} f\left(g\left(\e^{-\sigma}
\right)\right)} \nn
&&=  {A \over f'(A)} - {f(A) \over f'(A)^2}\ .
\eea
Especially, in case of \cite{CDTT}, when $A(=R)$ is small, $V(\sigma)$ behaves as a exponential function:
\be
\label{XXX12}
V(\sigma)\sim \frac{2}{\mu^2}\e^{\frac{2}{3}\sigma}\ .
\ee
Then we can solve the equations as in the previous section. 
Even for more general case \cite{NOgrg} as
\be
\label{XXX13}
f(A)=A + \gamma A^{-n}\left(\ln {A \over \mu^2}\right)^m\ ,
\ee
when curvature is small and $n\neq 0$, the potential behaves as an exponential function 
\be
\label{XXX14}
V(\sigma)\sim \left(1 + {1 \over n}\right) \left( -\gamma n\right)^{1 \over n+1}
\e^{{n+2 \over n+1}\sigma}\left({\sigma \over n+1}\right)^{-m} \ ,
\ee
and we can solve again the equations: 
\be
\label{XXX14b}
a\sim t^{(n+1)(2n+1) \over n+2}\ ,\quad w=-{6n^2 + 7n - 1 \over 3(n+1)(2n+1)}
\ee
Then we find that when
\bea
\label{XXX15}
&& n<\frac{-7 - 6\sqrt{2}}{12}\ ,\quad 
-1<n<-{1 \over 2}\ \nn
&& \mbox{or}\ n>{-7 + 6\sqrt{2} \over 12}
\eea
$w<0$ and when 
\be
\label{XXX16}
n< - \frac{1+\sqrt{3}}{2}\quad \mbox{or} \quad n> {-1 + \sqrt{3} \over 2}\ ,
\ee
the universe accelerates. 

In case that there is no matter and the Ricci tensor $R_{\mu\nu}$ is covariantly constant 
($\nabla_\rho R_{\mu\nu}=0$, which means 
$R_{\mu\nu}=\mbox{const.}\times g_{\mu\nu}$), the equation of motion is reduced to 
\be
\label{XXX17}
0=2f(R) - Rf'(R)\ ,
\ee
which is an algebraic equation with respect to $R$.
For example, when
\be
\label{XXX18}
f(R)=R - \frac{a}{R^n} + b R^m\ ,
\ee
we have
\be
\label{XXX19}
0=-R + \frac{(n+2)a}{R^n} 
+ (m-2)b R^m\ .
\ee
Especially, in case of \cite{CDTT} ($n=1$, $a=\mu^4$, $b=0$), we find
\be
\label{XXX20}
R=\pm \sqrt{3}\mu^2\ .
\ee
The $+$ sign corresponds to deSitter space, which is an exponentially expanding and accelerating universe. 
We may regard this solution would correspond to the present universe.
We should not that even if $b\neq 0$ but $m=2$, we obtain the same solution as (\ref{XXX20}). 

Just after the $1/R$-model was proposed, there were several criticism \cite{DK,Chiba} and an 
improvement was proposed in \cite{NOpn}, where the action is modified as 
\be
\label{XXX21}
f(R)=R - \frac{a}{R} + bR^2\ .
\ee
Simultaneously with the acceleration of the present universe, the model may explain the inflation 
at the early stage. The models of the inflation using $R^2$-theory has already 
been investigated in \cite{AAS,MMS}. We should note that when $R$ is large, where $n$ goes to $2$, 
from (\ref{XXX14}), we find $w\to -1$, which corresponds to the deSitter space. 

First, we consider the instability pointed out in \cite{DK}. 
The general equation of motion with matter in $f(R)$-gravity is given by
\bea
\label{XXX22}
&& {1 \over 2}g_{\mu\nu} f(R) - R_{\mu\nu} f'(R) - g_{\mu\nu} \Box f'(R) 
+ \nabla_\mu \nabla_\nu f'(R) \nn
&& = - {\kappa^2 \over 2}T_{(m)\mu\nu}\ .
\eea
Here $T_{(m)\mu\nu}$ is the matter energy-momentum tensor. By multipling the above equation with 
$g^{\mu\nu}$, we obtain
\bea
\label{XXX23}
&& \Box R + {f^{(3)}(R) \over f^{(2)}(R)}\nabla_\rho R \nabla^\rho R 
+ {f'(R) R \over 3f^{(2)}(R)} - {2f(R) \over 3 f^{(2)}(R)} \nn
&& = {\kappa^2 \over 6f^{(2)}(R)}T\ .
\eea
Here $T\equiv T_{(m)\rho}^{\ \rho}$
We consider a perturbation from the solution of the Einstein gravity:
\be
\label{XXX24}
R=R_0\equiv -{\kappa^2 \over 2}T>0\ .
\ee
We should note that $T$ is negative since $|p|\ll \rho$ in the earth and 
$T=-\rho + 3 p \sim -\rho$. Then we assume
\be
\label{XXX25}
R=R_0 + R_1\ ,\quad \left(\left|R_1\right|\ll \left|R_0\right|\right)\ .
\ee
Then we find
\bea
\label{XXX26}
0&=&\Box R_0 + {f^{(3)}(R_0) \over f^{(2)}(R_0)}\nabla_\rho R_0 \nabla^\rho R_0 
+ {f'(R_0) R_0 \over 3f^{(2)}(R_0)} \nn
&& - {2f(R_0) \over 3 f^{(2)}(R_0)} - {R_0 \over 3f^{(2)}(R_0)} \nn
&& + \Box R_1 + 2{f^{(3)}(R_0) \over f^{(2)}(R_0)}\nabla_\rho R_0 \nabla^\rho R_1 + U(R_0) R_1\nn
U(R_0)&\equiv& \left({f^{(4)}(R_0) \over f^{(2)}(R_0)} - {f^{(3)}(R_0)^2 
\over f^{(2)}(R_0)^2}\right) \nabla_\rho R_0 \nabla^\rho R_0 + {1 \over 3}R_0 \nn
&& - {f^{(1)}(R_0) f^{(3)}(R_0) R_0 \over 3 f^{(2)}(R_0)^2} 
 - {f^{(1)}(R_0) \over 3f^{(2)}(R_0)} \nn
&& + {2 f(R_0) f^{(3)}(R_0) \over 3 f^{(2)}(R_0)^2} 
 - {R_0 f^{(3)} \over 3f^{(2)}(R_0)^2} 
\eea
If $U(R_0)$ is positive, since $\Box R_1 \sim - \partial_t^2 R_1$, 
the perturbation $R_1$ becomes exponentially large and the system 
becomes unstable. We may regard $\nabla_\rho R_0\sim 0$ if we assume the matter is 
almost uniform as inside the earth.

In case of the model in \cite{CDTT}, we find 
\be
\label{XXX27}
U(R_0) = - R_0 + {R_0^3 \over 6\mu^4}
\ee
If $\mu$ corresponds to the acceleration of the present universe as in (\ref{XXX20}), we obtain
\be
\label{XXX28}
\mu^{-1}\equiv a^{-{1 \over 4}}\sim 10^{18} \mbox{sec} \sim \left( 10^{-33} \mbox{eV} \right)^{-1}\ .
\ee
This corresponds to the deSitter space which expands exponentialy. If the present universe expands with 
power law as in (\ref{XXX14}), the value of $a$ can be much larger than the above value.

The value of $R_0$ has been evaluated in \cite{DK}:
\bea
\label{XXX29}
{R_0^3 \over a}&\sim & \left(10^{-26} \mbox{sec}\right)^{-2} \left({\rho_m \over 
\mbox{g\,cm}^{-3}}\right)^3\ ,\nn 
R_0 &\sim&  \left(10^3 \mbox{sec}\right)^{-2} 
\left({\rho_m \over \mbox{g\,cm}^{-3}}\right)\ .
\eea
Then the system is unstable and would decay in $10^{-26}$ sec.

In the model (\ref{XXX21}) \cite{NOpn} with $a=\mu^4$, if $b\gg {a \over \left| R_0^3 \right|}$, 
we find 
\be
\label{XXX30}
U(R_0)\sim {R_0 \over 3}>0\ .
\ee
The system is unstable again but the decay time becomes about $1000$ second and macroscopic, that is, 
it is improved by $10^{29}$. But the assumption $b\gg {a \over \left| R_0^3 \right|}$ is not so realistic 
since this means $b^{-1}\ll \left(10^{11} {\rm eV}\right)^2=\left(10^{2} {\rm GeV}\right)^2$. 
In \cite{HN}, it has been given the constraint for inflation in $R^2$-theory from  COBE-DMR data of 
CMBR (cosmic microwave background radiation) as 
\be
\label{XXX31}
\frac{M}{m_{pl}} = \frac{b^{-\frac{1}{2}}}{m_{pl}} \sim 2.6\times 10^{-6}\ ,
\ee
that is $b^{-1} \sim \left(10^{12} {\rm GeV}\right)^2$, which contradicts with the assumption. 
Anyway instability depends on the details of $f(R)$. 
Furthermore, if the present universe expands by 
the power law, we need not to assume
\be
\label{XXX32} 
\mu^{-1}\equiv a^{-{1 \over 4}}\sim 10^{18} \mbox{sec} \sim \left( 10^{-33} \mbox{eV} 
\right)^{-1}\ .
\ee

In \cite{Chiba}, it has been pointed out the problem in the equivalence principle. 
By the scale transformation, the matter action is transformed as
\be
\label{XXX33}
S\left(g_{\mu\nu}, \psi\right) \to S\left(\e^\sigma g_{\mu\nu}, \psi\right) 
\ee
Then the matter couples with $\sigma$, which may violate the equivalence principle due to the 
force mediated by $\sigma$. 
In case of the model in \cite{CDTT}, if we assume (\ref{XXX32}) again, the mass of $\sigma$ 
becomes  $10^{-33}$ eV$=10^{-42}$ GeV and very light, which will violate the 
equivalence principle. 

In the model \cite{NOpn}, in the neighbourhood of $R=A=\sqrt{3a}=\sqrt{3}\mu^2$, we obtain
\bea
\label{XXX34}
\lefteqn{\left.{d^2 V(\sigma) \over d\sigma^2}\right|_{A=\sqrt{3a}}=
\left\{\left({d\sigma \over dA}\right)^{-2}
\left.{d^2 V(A) \over dA^2}\right\}\right|_{A=\sqrt{3a}}} \nn
&& ={ \sqrt{3a} \left({1 \over 3} + 2b\sqrt{3a}\right)^2 \over 
\left( - {1 \over 3} + b\sqrt{3a}\right)\left({4 \over 3} 
+ 2b\sqrt{3a}\right)^3} \ .
\eea
Then if
\be
\label{XXX35}
b \sim {1 \over 3 \sqrt{3a}}\ ,
\ee
which means $f'(A)\sim 0$, the mass of $\sigma$ becomes very large and would not conflict 
with the equivalence principle. 
In general, since
\bea
\label{XXX36}
V(\sigma)&=& \e^\sigma g\left(\e^{-\sigma}\right) - \e^{2\sigma} f\left(g\left(\e^{-\sigma}
\right)\right) \nn
&=& {A \over f'(A)} - {f(A) \over f'(A)^2}
\eea
the mass of $\sigma$ would become large when $f'(A)\sim 0$. Then if the present acceleration of universe 
corresponds to deSitter solution $R=A_0>0$, if we consider a model where $f'(A_0)\sim 0$, 
the mass of $\sigma$ can be large.  

\section{Big Rip Singularity}

In case of phantom ($w<-1$), the scale factor of the universe behaves as 
$a=a_0 \left(t_s - t\right)^\frac{2}{3(w+1)}$. Then $a$ diverges at $t\to t_s$. Then anything like atom or hadron, 
which has internal structure, should be torn. Then this singularity is called the Big Rip singularity 
(``rip'' meas a long tear or cut). This is the third possibility of the fate in the universe,  following 
the Big Crunch and the eternal exapnsion. 

We now consider the origin of the Big Rip. First we note that the energy density $\rho$ behaves as  
\be
\label{XXX37}
\rho=\rho_0 a^{-3(1+w)}\ .
\ee
Then in case of phantom ($w<-1$), when $a$ increases, $\rho$ also increases. 

In order to consider any concrete model, we consider the scalar field theory coupled with gravity, 
whose action is given by
\be
\label{XXX38}
S={1 \over \kappa^2}\int d^4 x \sqrt{-g} \left( R - {\gamma \over 2}
\partial_\mu \phi \partial^\mu \phi - V(\phi) \right)\ .
\ee
If $\gamma<0$, we have a phantom. By combining the equation given by the variation of $\phi$:
\be
\label{XXX39}
0=-\gamma \left(\frac{d^2 \phi}{dt^2} + 3H \frac{d\phi}{dt}\right) - V'(\phi)\ .
\ee
and the (1st) FRW equation : 
\be
\label{XXX40}
{6 \over \kappa^2}H^2=\rho_\phi={\gamma \over 2}\left(\frac{d\phi}{dt}\right)^2 + V(\phi)\ ,
\ee
we find
\be
\label{XXX41}
\frac{d\rho_\phi}{dt}= -3\gamma H \left(\frac{d\phi}{dt}\right)^2\ ,
\ee
which is positive if $\gamma<0$, $H>0$, and $\dot\phi\neq 0$. 
Then the energy density of phantom ($\gamma<0$) increases with time. 

We now consider if the Big Rip singularity really exist or not. 
First we should note that since $\gamma<0$, we have $\rho_\phi \leq V(\phi)$. 
Then if there is an upper bound in $V(\phi)$, there is also an upper bound in $\rho_\phi$. 
The potential  $V(\phi)$ might behave exponentially in the present epoch of the universe, 
but if there is an upper bound in $V(\phi)$, there does not occur the Big Rip singularity and 
the space becomes deSitter space asymptotically. 

Another possibility to avoid the Big Rip singularity is to include the quantum correction \cite{NOqc}
\footnote{
Phantom theories with quantum corrections have been discussed in \cite{NOphantom}. 
The possibility to avoid the Big Rip singularity by the quantum effects has been also considered 
in \cite{Onemli}. }. 
Near the Big Rip singularity, since $a$ blows up, cuvatures become large as $R\propto \left|t - t_s\right|^{-2}$. 
Since the quantum correction contains power of the curvatures in general, the quantum correction becomes 
very important. 
In order to consider such a quantum correction, we include the contribution from conformal anomaly :
\be
\label{XXX42}
T=b\left(F+{2 \over 3}\Box R\right) + b' G + b''\Box R\ ,
\ee
Here $F$ is the square of 4d Weyl tensor and $G$ is the  Gauss-Bonnet invariant:
\bea
\label{XXX43}
F&=&{1 \over 3}R^2 -2 R_{ij}R^{ij}+ R_{ijkl}R^{ijkl} \ , \nn
G&=&R^2 -4 R_{ij}R^{ij}+ R_{ijkl}R^{ijkl} \ ,
\eea
In case that there are $N$ scalar, $N_{1/2}$ Dirac spinor, $N_1$ vector fields, $N_2$ ($=0$ or $1$) 
gravitons, $N_{\rm HD}$ higher derivative conformal scalars, the coefficients $b$, $'$, and $b''$ 
are given by
\bea
\label{XXX44}
b&=&{N +6N_{1/2}+12N_1 + 611 N_2 - 8N_{\rm HD} \over 120(4\pi)^2}\ ,\nn
b'&=&-{N+11N_{1/2}+62N_1 + 1411 N_2 -28 N_{\rm HD} \over 360(4\pi)^2}\ , \nn
b''&=&0\ .
\eea
We should note $b>0$ and $b'<0$ for the usual matter except the higher derivative conformal scalars.
We should note that $b''$ can be shifted by the finite renormalization of the local counterterm $R^2$. 
Then $b''$ can be arbitrary. 
By using the corresponding  energy density $\rho_A$ and pressure $p_A$, 
$T_A$ is given by $T_A=-\rho_A + 3p_A$. 
Then by using the energy conservation in the FRW universe
\be
\label{CB1}
0=\frac{d\rho_A}{dt} + 3 H\left(\rho_A + p_A\right)\ ,
\ee
we may delete $p_A$ as 
\be
\label{CB2}
T_A=-4\rho_A - \frac{1}{H}\frac{d\rho_A}{dt}\ ,
\ee
which gives the following expression of $\rho_A$:
\bea
\label{CB3}
\rho_A&=& -\frac{1}{a^4} \int dt a^4 H T_A \nn
&=&  -\frac{1}{a^4} \int dt a^4 H \left[ -12b \left(\frac{dH}{dt}\right)^2 \right. \nn
&& + 24b'\left\{  - \left(\frac{dH}{dt}\right)^2 + H^2 \frac{dH}{dt} + H^4\right\} \nn
&& - 6\left(\frac{2}{3}b + b''\right)\left\{ \frac{d^3H}{dt^3} + 7 H \frac{d^2 H}{dt^2} \right. \nn
&& \left. \left. + 4 \left(\frac{dH}{dt}\right)^2 + 12 H^2 \frac{dH}{dt}\right\}\right]\ .
\eea
Then we also have an expression for $p_A$
\bea
\label{CB4}
p_A&=& - \rho_A - \frac{1}{3H}\frac{d\rho_A}{dt} \nn
&=& \frac{T_A}{3}  -\frac{1}{a^4} \int dt a^4 H T_A \nn
&=&  \frac{1}{3}\left[ -12b \left(\frac{dH}{dt}\right)^2 \right. \nn
&&+ 24b'\left\{  - \left(\frac{dH}{dt}\right)^2 + H^2 \frac{dH}{dt} + H^4\right\} \nn
&& -  6\left(\frac{2}{3}b + b''\right)\left\{ \frac{d^3H}{dt^3} + 7 H \frac{d^2 H}{dt^2} \right. \nn
&& \left.\left. + 4 \left(\frac{dH}{dt}\right)^2 + 12 H^2 \frac{dH}{dt}\right\}\right] \nn
&& -\frac{1}{a^4} \int dt a^4 H \left[ -12b \left(\frac{dH}{dt}\right)^2 \right. \nn
&& + 24b'\left\{  - \left(\frac{dH}{dt}\right)^2 + H^2 \frac{dH}{dt} + H^4\right\} \nn
&& - 6\left(\frac{2}{3}b + b''\right)\left\{ \frac{d^3H}{dt^3} + 7 H \frac{d^2 H}{dt^2} \right. \nn
&& \left.\left. + 4 \left(\frac{dH}{dt}\right)^2 + 12 H^2 \frac{dH}{dt}\right\}\right]\ .
\eea
By including the quantum correction in (\ref{CB3}), we may write the corrected FRW equation in 
the following form:
\be
\label{BR7}
{6 \over \kappa^2}H^2 ={\gamma \over 2}\left({d\phi \over dt}\right)^2 + V(\phi) + \rho_A\ .
\ee
We now assume
\be
\label{BR8}
H=h_0 + \delta h \ ,\quad \phi=\phi_0\ln \left|\frac{t_s - t}{t_1}\right| + \delta\phi\ .
\ee
and when $t\to t_s$, $\delta h$, $\delta\phi$ are much smaller than the first terms but 
$\frac{d\delta h}{dt}$ might be singular. 
Then $\phi$-equation of motion reduces as 
\begin{eqnarray}
\label{BR9}
\lefteqn{0=-\gamma\left( - \frac{\phi_0}{\left(t_s - t\right)^2} - \frac{3h_0}{t_s - t}\right)} \nn
&& + \frac{2V_0t_1^2}{\phi_0\left(t_s - t\right)^2}\left(1 - \frac{2}{\phi_0}\delta\phi\right) 
+ o\left(\left(t_s - t\right)^{-1}\right)\ ,
\end{eqnarray}
which gives 
\be
\label{BR10}
V_0t_1^2= - {\gamma\phi_0^2 \over 2}\ ,\quad \delta\phi=-\frac{3}{2}\left(t_s - t\right)\ .
\ee
If we assume $\frac{2}{3}b + b''\neq 0$, we find
\be
\label{BR11}
\rho_A \sim 6h_0 \left(\frac{2}{3}b + b''\right)
\frac{d^2 \delta h}{dt^2}\ .
\ee
Then by substituting (\ref{BR10}) and (\ref{BR11}) into the corrected FRW equation (\ref{BR7}), 
we find  
\bea
\label{BR12}
0&=&\frac{3\gamma h_0 \phi_0}{t_s - t} + 6h_0 \left(\frac{2}{3}b + b''\right)
\frac{d^2 \delta h}{dt^2} \nn
&& + o\left(\left(t_s - t\right)^{-1}\right)\ ,
\eea
and 
\be
\label{BR13}
\delta h=\frac{\gamma\phi_0}{2 \left(\frac{2}{3}b + b''\right)}\left(t_s - t\right)
\ln \left|\frac{t_s - t}{t_2}\right| \ .
\ee
Here $t_2$ is a constant of the integration. 
Then the scale factor $a$ behaves as
\bea
\label{BR14}
a&=&a_0\left|\frac{t_s - t}{t_2}\right|^{\frac{\gamma\phi_0}{4 \left(\frac{2}{3}b + b''\right)}
\left(t_s - t\right)^2} \nn
&& \times \e^{-h_0\left(t_s - t\right) 
 - \frac{\gamma\phi_0}{8 \left(\frac{2}{3}b + b''\right)}\left(t_s - t\right)^2
+ o\left(\left(t_s - t\right)^2\right)}\ .
\eea
There appear logarithmic singularities in $\frac{d^2 a}{dt^2}$, $\frac{dH}{dt}$ but the singularity 
becomes rather mild. Then the universe might develope beyond $t=t_s$. 

\section{Summary and Discussion.}

In summary, there are several models to explain the accelerated expansion of the present universe. 
For the Big Rip singularity, the quantum correction would be very important. 

There are many other models, as tachyon (see \cite{NOt} for an example),  
the Born-Infeld theory (see \cite{ELNO} for example), Chameleon model \cite{Chame} 
(see also \cite{NOchame}), and $R+L_{\rm matter}R^n$ type model in \cite{NOlr}. 
There are applications of the modified gravities to the cosmology (for example, \cite{Zhu}). 
These theories have structures as in the models obtained by the dimensional reduction 
from the model in the higher dimension. 
In fact, several models are proposed from the viewpoint of the brane world
(for example, see \cite{Deffayet}). 
Near $w=-1$, we feel that theories in the extra or higher dimensions (brane world, Kaluza-Klein model) 
would appear.

\section*{Acknowledgments}

The author is indebted with S.D. Odintsov for the collaborations. 
This investigation has been supported in part by the
Ministry of Education, Science, Sports and Culture of Japan under
grant n.13135208.

\end{document}